\documentclass[10pt,conference]{IEEEtran}
\IEEEoverridecommandlockouts
\usepackage{cite}
\usepackage{amsmath,amssymb,amsfonts}
\usepackage{graphicx}
\usepackage{textcomp}
\usepackage{xcolor}
\usepackage{dcolumn} 
\usepackage{bm}      
\usepackage{hyperref} 
\usepackage{mathtools} 
\usepackage{algorithm}
\usepackage{algpseudocode}
\usepackage{braket}
\usepackage{booktabs}
\usepackage{quantikz}

\begin{document}

\title{A New Algorithm for Applying Sequences of Affine Transformations in Quantum Circuits%
\thanks{This work was supported by the National Science Foundation under Grant Number 2442853.}}

\author{
\IEEEauthorblockN{Anish Giri}
\IEEEauthorblockA{Department of Computer Science\\
Vanderbilt University\\
Nashville, Tennessee 37212, USA\\
Email: anish.giri@vanderbilt.edu}
\and
\IEEEauthorblockN{David Hyde,~\IEEEmembership{Senior Member,~IEEE}}
\IEEEauthorblockA{Department of Computer Science\\
Vanderbilt University\\
Nashville, Tennessee 37212, USA\\
Email: david.hyde.1@vanderbilt.edu}
\and
\IEEEauthorblockN{K\'alm\'an Varga}
\IEEEauthorblockA{Department of Physics and Astronomy\\
Vanderbilt University\\
Nashville, Tennessee 37240, USA\\
Email: kalman.varga@vanderbilt.edu}
}

\maketitle

\begin{abstract}
This paper introduces a novel technique for applying sequences of affine transformations in quantum circuits.
Utilizing Hadamard-supported conditional initialization and block encoding, the proposed method systematically applies sequential affine transformations while preserving state normalization.
Interleaved amplitude amplification ensures that long sequences of affine transformations can be applied without exponential amplitude decay of the desired state.
We provide preliminary demonstrations of our algorithm on a simple discrete signal processing task and on a driven Hamiltonian dynamics example.
\end{abstract}


\section{Introduction}
\label{sec:intro}
Affine transformations combine linear transformations and translations as \( A\Psi + B \), where \( \Psi \in \mathbb{R}^N \) is the input vector, \( A \in \mathbb{R}^{N \times N} \) is the linear transformation matrix, and \( B \in \mathbb{R}^N \) is the translation vector. 
Affine transformations are foundational in machine learning \cite{A1} and iterative methods \cite{A2} for tasks like image processing, regression, and data analysis, where the relative positions of elements need to be retained while allowing for scaling, rotations, and shear transformations.

In quantum computing, the implementation of affine transformations presents a challenge because quantum operations are inherently unitary (time reversible), i.e., \( U^\dagger U = I \), preserving the norm of quantum states.
However, a simple affine transformation will alter the norm of a state vector, which would suggest that one cannot implement affine transformations with quantum circuits.
However, recent works have demonstrated multiple approaches for quantum implementations of affine transformations. 
For instance, one approach proposes encoding pixel positions and employing controlled-NOT (CNOT) gates and adders for operations like scaling, translation, and scrambling, thereby enabling efficient manipulation of quantum-encoded images \cite{paper1}.
Another approach leverages the synthesis of reversible circuits from invertible matrices, using CNOT gates to represent linear transformations and combining them with simple classical elements like NOT and XOR gates for affine shifts \cite{paper2}.
From a more structural perspective, polyhedral models have been utilized to optimize the scheduling and allocation of gates in circuits that apply affine transformations, effectively handling execution dependencies through loop transformations and affine maps \cite{paper3}.
Polyhedral abstractions combined with integer linear programming (ILP) techniques have enabled the derivation of optimal qubit indexing schemes, minimizing gate counts and circuit depths while respecting device constraints \cite{paper5}.
Quantum circuits that realize affine mappings over finite fields \( F_{2^n} \) have been explored using modular arithmetic and normal basis representations, yielding efficient and scalable arithmetic circuits \cite{paper4}. Similarly, Linear Combination of Unitaries (LCUs) are often used to deal with unnormalized quantum states localized in a subspace of ancilla qubit(s), where junk branch's amplitudes are neglected \cite{chakraborty2024implementing}.

While these works illustrate the potential of implementing affine transformations in quantum settings, there remains a need for a more scalable algorithm that can apply a series of affine transformations.
One traditional approach, which borrows from classical fields like computer graphics \cite{marschner2016fundamentals}, uses homogeneous coordinates to define an augmented state vector on which the affine transformation \( f(\Psi) = A\Psi + B \) can be performed using a single matrix-vector multiplication in a higher-dimensional space.
Specifically, we may define 
\begin{equation*}
\tilde{\Psi}
=
\frac{1}{\sqrt{2}}
\begin{bmatrix}
\Psi^{T} & 0 & 0 & \cdots & 1
\end{bmatrix}^{T}
\in \mathbb{R}^{2N}.
\end{equation*}
and then the augmented matrix
\begin{equation*}
\tilde{A} = 
\begin{bmatrix} 
A^{N \times N} & 0^{N \times (N-1)} \quad B^{N \times 1} \\[1ex]
0^{N \times N} & I^{N \times N} 
\end{bmatrix} 
\in \mathbb{R}^{(2N) \times (2N)}.
\end{equation*}
With this representation, the affine transformation can be expressed as a purely linear transformation in the augmented space:
$
\tilde{f}(\tilde{\Psi}) = \tilde{A} \tilde{\Psi}.
$
However, this matrix is not unitary.
To correct this, we can perform a standard block encoding trick, which embeds the \( (2N) \times (2N) \) normalized non-unitary matrix \( \tilde{A}/\alpha \) into a larger \( (4N) \times (4N) \) unitary matrix
\[
U(\tilde{A}) = 
\begin{pmatrix}
\tilde{A}/\alpha & \sqrt{I - \tilde{A}\tilde{A}^\dagger/\alpha^2} \\
\sqrt{I - \tilde{A}^\dagger \tilde{A}/\alpha^2} & -\tilde{A}^\dagger/\alpha
\end{pmatrix},
\]
where \( \alpha > 0 \) is a normalization factor to ensure that \( \|\tilde{A} / \alpha\| \leq 1 \) \cite{E1,E2,Ad2}. The encoded unitary $U(\tilde{A})$ can be applied conditionally to the desired number of qubits to produce the affine-transformed amplitudes in the quantum state.
In this way, block encoding can be combined with linear transformations in augmented spaces to achieve affine transformations of vectors or fields.

However, constructing a block-encoding procedure is challenging and becomes increasingly computationally demanding as the matrix size grows. Employing an augmented affine matrix using block encoding leads to a fourfold increase in matrix size, which can necessitate a greater number of qubits and makes decomposing the operator into a product of basis gates more challenging \cite{E3,E4,E6} (at this time, factoring a unitary into an approximate product of basis gates remains challenging even for a small number of qubits \cite{2408.15225}). Finally, it is even harder to create a quantum circuit from a single augmented affine matrix using block-encoding in cases like driven dynamics evolution demonstrated in the applications section \ref{sec:apps}. These considerations motivate the search for more scalable and efficient strategies to perform affine transformations within the memory of a quantum computer.

In this paper, we introduce a scalable method for applying sequences of affine transformations on a field localized within a sub-state of a quantum system. 
Our approach uses block encoding of \( A \) instead of \(\tilde{A}\), which leads to a matrix size of $4N^2$ instead of $16N^2$, and employs Hadamard-supported conditional initialization of the translation terms.
This method not only facilitates execution of sequences of affine transformations but also generates combinatorial patterns of additive and subtractive translation elements, effectively accumulating these amplitudes across various sub-states.
This holds possibilities for further quantum computing applications such as quantum combinatorial optimization, signal processing, and quantum machine learning \cite{Ad1,Ad3,Ad4}.
With recent advancements in quantum computing algorithms for solving differential equations \cite{C1,C2,C5}, combinatorial optimization through quantum gradient descent \cite{C4}, quantum approaches to solving solid mechanics problems \cite{C3}, and plasma simulations \cite{C6}, an algorithm of this nature can potentially offer efficient iterative methods for addressing these complex problems. \\
The paper is organized as follows.
In the next section, we present a simple method for element-wise addition and subtraction in an \( n+1 \)-qubit system.
Following this, a step-by-step method for applying sequences of affine transformations is introduced.
The result is generalized to any number of affine transformations for any number of qubits.
We then discuss how interleaving amplitude amplification operations into the algorithm guards against decay of the amplitude of the desired state.
Finally, we show simple yet representative examples of our algorithm on a discrete signal processing task and on a driven dynamics evolution problem.

\section{Hadamard-supported Element-wise Addition and Subtraction}
\label{sec:addsub}
Several approaches have been developed to achieve arithmetic operations within quantum states using Fourier-arithmetic techniques \cite{B1,B2,B3,B4}. 
However, these methods are either not suitable or applicable for performing arithmetic directly on a large number of field values within a single quantum state while allowing the flexibility to apply extra unitary operations within the quantum circuit.
Instead, we build our idea on using amplitudes of quantum states to represent the field values and perform arithmetic on those amplitudes.

A significant challenge in quantum systems is the inability to directly add or subtract bias terms from amplitudes while maintaining normalization.
For instance, consider the states:
\begin{equation*}
\Psi_A = a_1\ket{00} + a_2\ket{01} + a_3\ket{10} + a_4\ket{11}, 
\end{equation*}
\begin{equation*}
\Psi_B = b_1\ket{00} + b_2\ket{01} + b_3\ket{10} + b_4\ket{11}.
\end{equation*}
If we simply added the two sets of amplitudes, we could end up with a state whose total probability (the $L_2$ norm of the vector of probability amplitudes) exceeds 1:
\begin{equation*}
\begin{split}
\Psi_{(A + B)} &= (a_1 + b_1)\ket{00} + (a_2 + b_2)\ket{01} \\
              &\quad + (a_3 + b_3)\ket{10} + (a_4 + b_4)\ket{11}.
\end{split}
\end{equation*}

However, it is possible to create a quantum state where \( a_i + b_i \) is accumulated over states, scaled by a known factor, using Hadamard-supported conditional initialization as demonstrated in Fig.~\ref{fig:add_sub}.
We note that conditional initialization, in general, is often available in quantum computing software development kits, such as Qiskit; and a similar approach in which \( \Psi_A \) and \( \Psi_B \) are sub-states of a quantum state and are conditionally implemented between two Hadamard gates on the ancillary qubit is demonstrated in \cite{D1}, which is a similar process to using LCUs.
Conditional initialization prepares one sub-state when the ancilla is \( \ket{0} \) and another sub-state when the ancilla is \( \ket{1} \). Suppose the two quantum states are given by:
$
\Psi_A = \sum_{i=1}^{N} a_i \ket{i}, \quad \Psi_B = \sum_{i=1}^{N} b_i \ket{i}, 
$
where $N = 2^n$.
The system is prepared with an \((n+1)\)-qubit register in the state:
$
\Psi_0 = \ket{0} \otimes \ket{0}^{\otimes n}.
$
Here, the first qubit acts as an ancillary qubit, while the remaining \( n \) qubits are the target register that will store the quantum states. A Hadamard gate \( H \) is applied to the ancilla, creating an equal superposition of its computational basis states:
\begin{equation*}
\Psi_1 = \frac{1}{\sqrt{2}} (\ket{0} + \ket{1}) \otimes \ket{0}^{\otimes n}.
\end{equation*}
This superposition ensures that subsequent initialization of the target qubits depends conditionally on the ancilla's state. Conditional initialization is then applied to the last \( n \) qubits. When the ancilla is in \( \ket{0} \), the target qubits are prepared in the state \( \Psi_B \). When the ancilla is in \( \ket{1} \), the target qubits are prepared in the state \( \Psi_A \). The resulting state becomes:
\begin{equation*}
\Psi_2 = \frac{1}{\sqrt{2}} \left( \ket{0} \otimes \Psi_B + \ket{1} \otimes \Psi_A \right). 
\end{equation*}
Finally, a second Hadamard gate is applied to the ancilla, mixing the states and producing:
\begin{equation*}
\Psi_3 = \frac{1}{2} \left[ \ket{0} \otimes (\Psi_A + \Psi_B) + \ket{1} \otimes (\Psi_A - \Psi_B) \right].
\end{equation*}
Expanding this, the amplitudes of \( \Psi_A \) and \( \Psi_B \) are combined:
\begin{equation*}
\Psi_3 = \frac{1}{2} \sum_{i=1}^{N} \Bigl[ (a_i + b_i) \ket{0}\ket{i} + (a_i - b_i) \ket{1}\ket{i} \Bigr]. 
\end{equation*}
This final state encodes both the element-wise sum \( a_i + b_i \) and difference \( a_i - b_i \), scaled to preserve normalization. The final state can yield just the scaled sum or just the scaled difference by measuring the ancilla qubit in the $\{|0\rangle, |1\rangle\}$ basis and then projecting the state onto either the sum ($|0\rangle$) or the difference ($|1\rangle$) component.

\begin{figure}[htbp]
\centering
\[
\begin{quantikz}[row sep={0.4cm,between origins}, column sep=0.6cm]
\lstick{$\ket{0}$} & \gate{H} & \octrl{1}      & \ctrl{1}      & \gate{H} & \qw \\
\lstick{$\ket{0}$} & \qw      & \gate[2]{\Psi_B} & \gate[2]{\Psi_A} & \qw      & \qw \\
\lstick{$\ket{0}$} & \qw      & \ghost{\Psi_B} & \ghost{\Psi_A} & \qw      & \qw
\end{quantikz}
\]
\caption{Hadamard-supported element-wise addition and subtraction for a three-qubit system, where the first qubit is ancillary and the remaining two qubits are conditionally initialized with \( \Psi_A \) and \( \Psi_B \).}
\label{fig:add_sub}
\end{figure}

We note that implementations of conditional state initialization are enabled by recent advancements in quantum state initialization algorithms, particularly the divide-and-conquer and top-down approaches.
The divide-and-conquer algorithm achieves an exponential reduction in circuit depth, scaling from \( O(N) \) in traditional top-down methods to \( O(\log_2^2(N)) \) \cite{F1,F2}.
However, this improvement comes at the cost of increased qubit usage, requiring \( O(N) \) ancillary qubits compared to \( O(\log_2(N)) \) qubits in top-down methods \cite{F4}.
Depending on the requirements and resource constraints, either state initialization technique can be applied conditionally.

\section{Sequential Affine Transformations}
\label{sec:sat}
In this section, we discuss how to apply sequential affine transformations to an input field that is accumulated over a sub-state of a quantum system. Given a discrete field \( \Psi \in \mathbb{R}^N \), the objective is to achieve the following series of transformations:
\[
\Psi_{\text{final}}^{N} = A^{N \times N}_k \big( \cdots A^{N \times N}_2 \big(A^{N \times N}_1 \Psi + B^N_1 \big) + B^N_2 \big) \cdots + B^N_k
\]
Here, \( A^{N \times N}_j \) is an \( N \times N \) linear transformation matrix, and \( B^N_j \) is a column translation vector applied at each step \( j \).

To achieve the final amplitudes for this transformation, accumulated over a sub-state in a quantum circuit, the circuit width must be strategically expanded, along with constructing a unitary version of each \( A_j \).
Additionally, each \( B_j \) must be resized appropriately to match the expanded dimensions of the associated unitary version of \( A_j \). 
At each step \( j \), the process involves Hadamard-supported element-wise addition and subtraction, followed by conditional initialization of the resulting state with an additional ancillary qubit.
These steps ensure the sequential application of the affine transformations within the quantum circuit. The complete sequence of operations is outlined as follows:

\vspace{0.5em}
\noindent \textbf{Step 1: Amplitude Initialization}
Starting with an \( s \)-qubit system initialized in the state \( \ket{0}^{\otimes s} \), 
we encode a given (user-supplied) normalized vector $\Psi = [x_1, x_2, \dots, x_N]^T \in \mathbb{R}^N$ (or can be complex, look into the footnote) into the amplitudes of the last \( n \) qubits ($n \leq s$) of the system, where \( N = 2^n \)\footnote{Both our method and the technique on the first page of the manuscript assume a normalized input state. It is possible to extend either technique to handle non-normalized input states with the addition of ancillary qubits and with larger matrices. Similarly, we can also use complex amplitudes, which creates extra cost on tomography to retrieve exact amplitudes.}.
This yields an initialized state
\[
\Psi_0 = \ket{0}^{\otimes (s-n)} \otimes \Bigl( \sum_{i=1}^{N} x_i \ket{i} \Bigr).
\]
\noindent \textbf{Step 2: Apply Unitary Transformation}
To apply the first transformation matrix \( A_1^{N\times N} \), we construct a unitary \( U_{1} \in \mathbb{R}^{(2N) \times (2N)} \) using block encoding. This unitary acts on \( n+1 \) qubits and transforms the state such that the amplitude corresponding to \( \ket{i} \) becomes \( x_i^{U_1} \). The resulting quantum state is given by:
\[
U_{1} \Psi_0 = \quad \ket{0}^{\otimes (s-n-1)} \otimes \Bigl( \sum_{i=1}^{2N} x_i^{U_1} \ket{i} \Bigr).
\]
Note that in the above we define $x_i = 0$ for $i= N+1, \dots, 2N$ (though by the action of $U_1$, the corresponding $x_i^{U_1}$ may be non-zero).
At this point, we have effectively applied $A_1^{N \times N}$ to our system by using an extra qubit, and the remaining step is to add $B_1^N$.

\vspace{0.5em}
\noindent \textbf{Step 3: Hadamard-supported Add/Sub with \( B_1^{N} \)} 
First, we zero-pad $B_1^{N}$ (which we also assume is normalized) to $B_1 \in \mathbb{R}^{2N}$ (i.e., extending to $n + 1$ qubits).
We apply Hadamard-supported addition and subtraction of
$B_1 = [\omega_1, \omega_2, \dots, \omega_{2N}]^T$
on the existing quantum states to perform amplitude arithmetic.
As suggested by Section \ref{sec:addsub}, this yields the state
\begin{equation*}
\begin{split}
\Psi_1 = \quad &\ket{0}^{\otimes (s-n-2)} \otimes \frac{1}{2} \Biggl( 
\sum_{i=1}^{2N} \Bigl( x_i^{U_1} + \omega_i \Bigr) \ket{0} \otimes \ket{i} \\
&\quad + \sum_{i=1}^{2N} \Bigl( x_i^{U_1} - \omega_i \Bigr) \ket{1} \otimes \ket{i} 
\Biggr).
\end{split}
\end{equation*}

After this step, the first affine transformation is achieved with its result---stored in the probability amplitudes of the first $N$ $|0i\rangle$ states, in the last $n+2$ qubits---scaled by a known factor of 1/2 (which can be undone classically by the user after measurement).
For clarity, we note that the $\ket{0} \otimes \ket{i}$ and $\ket{1} \otimes \ket{i}$ terms are inside their respective sums, not outside the sums.

\vspace{0.5em}
\noindent \textbf{Next Iteration:}
To establish the general pattern for applying sequences of affine transformations, we show what applying a second affine transformation then involves.
For the second affine transformation, we block encode \( A_2 \) into the unitary \( U_{2} \).
This unitary is applied to the last \( n+3 \) qubits of the quantum state as follows:
\begin{equation*}
\begin{split}
U_{2} \Psi_1 = \quad &\ket{0}^{\otimes (s-n-3)} \otimes \frac{1}{2} \Biggl( 
\sum_{i=1}^{4N} \Bigl( (x_i^{U_1} + \omega_i)^{U_2} \Bigr) \ket{0i} \\
&\quad + \sum_{i=1}^{4N} \Bigl( (x_i^{U_1} - \omega_i)^{U_2} \Bigr) \ket{1i} 
\Biggr).
\end{split}
\end{equation*}
We zero-pad the $x_i$ and $\omega_i$ as needed to make the above equation valid.

We then perform Hadamard-supported addition and subtraction using the normalized vector \( B_2^N \).
We wish to construct a $B_2$ analogous to $B_1$, but here we encounter a subtle detail where care is required.
We form the zero-padded vector $B_2 = [\beta_1, \beta_2, \dots, \beta_{8N}]^T$ ($\beta_i = 0$ for $i=N+1,\dots,8N$).
We then rescale the vector so that the first $4N$ entries become $\beta_i/2$, and the remaining entries are chosen via an appropriate scheme to ensure the state remains normalized\footnote{For instance, the second $4N$ entries ($i=4N+1,\dots,8N$), rather than being zero, could be set to $\sqrt{1-\beta_{i-4N}^2/4}$. Other strategies are possible.}.
We label this rescaled vector $\tilde B_2$.

With this $\tilde B_2$ in mind, applying Hadamard-supported additional and subtraction with $\tilde B_2$ modifies the amplitudes to yield a final state:
\begin{equation*}
\begin{split}
\Psi_2 = \quad &\ket{0}^{\otimes (s-n-4)} \otimes \frac{1}{2} \Biggl[ 
\sum_{i=1}^{8N} \Bigl( \frac{{(x_i^{U_1} + \omega_i)}^{U_2}}{2} + \tilde \beta_i \Bigr) \ket{0} \otimes \ket{0i} \\
&\quad + \sum_{i=1}^{8N} \Bigl( \frac{({(x_i^{U_1} - \omega_i)}^{U_2}}{2} + \tilde \beta_{i} \Bigr) \ket{0} \otimes \ket{1i} \\
&\quad + \sum_{i=1}^{8N} \Bigl( \frac{({(x_i^{U_1} + \omega_i)}^{U_2}}{2} - \tilde \beta_i \Bigr) \ket{1} \otimes \ket{0i} \\
&\quad + \sum_{i=1}^{8N} \Bigl( \frac{({(x_i^{U_1} - \omega_i)}^{U_2}}{2} - \tilde \beta_{i} \Bigr) \ket{1} \otimes \ket{1i}
\Biggr],
\end{split}
\end{equation*}
where $\tilde \beta_i$ is the $i$\textsuperscript{th} entry of $\tilde B_2$.
Note that for the first $2N$ terms in the first sum in the above expression, $\tilde \beta_i = \beta_i/2$.
So if one measures the first $N$ states $\ket{0} \otimes \ket{0i}$, $i = 1, \dots, N$, one can account for a known constant factor of $1/4$ that can be pulled out front, and one is thereby able to recover the result of the first and second affine transformation being sequentially applied.
We note that the remaining $7N$ states may hold useless values, which is irrelevant since we do not need to measure those states.

\section{Generalized Affine Transformation Steps}
\label{sec:generalized}

The generalized procedure for \(k\) affine transformation steps is described by a sequence of block-encoded unitaries and Hadamard-supported addition/subtraction operations.
At each step \(j\) (with \(1 \leq j \leq k\)), we have a linear transformation \(A_j^{N\times N}\) embedded into a unitary \(U_j\), which doubles the state dimension from \(2^{j-1}N\) to \(2^jN\), and a translation (or bias) vector \(B_j^N\) that is resized and normalized to form an associated \(\tilde{B}_j\).
Then a Hadamard-supported operation is applied that splits the state into two branches, labeled by a binary variable \(b_j \in \{0,1\}\), thus giving the amplitudes \[ \Phi^{(j)}_{(b_1,\dots,b_j),i} = \frac{1}{2}\Bigl[\Bigl(\Phi^{(j-1)}_{(b_1,\dots,b_{j-1}),i}\Bigr)^{U_j} + (-1)^{b_j}\,\tilde{\beta}^{(j)}_i\Bigr]. \]

Thus, if we start with the initial state $\Phi^{(0)}_i = x_i$ ($i=1,\dots,N$, with $N=2^n$) and have our associated normalized input state
\[
\Psi = \ket{0}^{\otimes (s-n)} \otimes \sum_{i=1}^{N} x_i\, \ket{i} ,
\]
then after \(k\) steps, the full quantum state is given by
\[
\Psi_{\text{final}} = \ket{0}^{\otimes (s-n-2k)} \otimes \Phi_{\text{aff}},
\]
\[
\Phi_{\text{aff}} = \frac{1}{2^k}\sum_{(b_1,\dots,b_k) \in \{0,1\}^k} \sum_{i=1}^{2^k N} \Phi^{(k)}_{(b_1,\dots,b_k),\,i}\, \ket{b_1b_2\cdots b_k,\,i}.
\]




\section{Circuit Construction}
So far, we have only seen how to scale up the affine transformation steps mathematically while preserving the norm. To transform this algorithm into a quantum circuit, we first fix the number of affine steps $k$. Say $k=3$. Then $\Psi_{final} = A_3(A_2(A_1 \Psi_{initial} + B_1) + B_2) + B_3 = A_3 A_2 A_1 \psi_{initial} + A_3 A_2 B1 + A_3 B_2 + B_3$. For $k$ steps of affine transformations, we need $k$ or $2k$ ancilla qubits, depending on whether each $A_i$ is unitary. Then these unitaries and initialization blocks can be applied in multi-control settings depending on
whether the ancilla qubits are in $\ket{0}$ or $\ket{1}$. An exact quantum circuit with $k=3$ and unitary $A_i$'s is demonstrated in the applications
section, Fig.~\ref{fig:driven_affine_circuit}. Once the total number of affine steps $k$ is fixed, the circuit can be built step by step: the first block implements the first affine transformation, the second block implements the second affine transformation, and so on up to the $k$-th block. For this fixed value of $k$, each additional affine step has approximately the same circuit cost, so the total implementation cost grows
linearly with the number of executed steps up to the fixed limit $k$.

\section{Interleaved Amplitude Amplification}
\label{sec:interleaved_aa}

The central scalability bottleneck of the sequential affine transformation
method is the $1/2^k$ amplitude decay arising from $k$ successive
Hadamard-supported conditional initializations. After $k$ steps, the target
sub-state is encoded with probability $1/4^k$, requiring either $O(4^k)$
measurement repetitions or $O(2^k)$ rounds of end-stage amplitude
amplification. We now show that \emph{interleaved} amplitude amplification, i.e. applying a single Grover iterate after each affine step, reduces the
total overhead from exponential to linear in $k$.

\subsubsection{Single-Step State Structure}
\label{sec:single_step_structure}

After the $j$-th affine transformation (block encoding of $U_j$ followed by
Hadamard-supported addition of $\tilde{B}_j$), the quantum state on the
full register has the form
\begin{equation*}
  \ket{\Phi^{(j)}} = \frac{1}{2}\ket{0}_{\mathrm{br}}
  \ket{0^a}_{\mathrm{anc}}\ket{\psi_j}_{\mathrm{data}}
  + \ket{\Psi_\perp},
\end{equation*}
where $\ket{0}_{\mathrm{br}}$ is the branching qubit introduced by the
Hadamard gate, $\ket{0^a}_{\mathrm{anc}} \equiv \ket{0}^{\otimes a}$
denotes the block-encoding ancilla register in the all-zero state,
$\ket{\psi_j}_{\mathrm{data}}$ is the $n$-qubit register containing the
desired affine-transformed amplitudes, and $\ket{\Psi_\perp}$ lies
entirely in the orthogonal complement of
$\ket{0}_{\mathrm{br}}\ket{0^a}_{\mathrm{anc}}$. Crucially, the target
component has amplitude exactly $1/2$ at every step, independent of $j$.

\subsubsection{Good-Subspace Oracle and Amplification}
To amplify the desired data branch, we define the good subspace as the one for
which the branching qubit and all block-encoding ancillas are in the zero
state, irrespective of the data register:
\begin{equation*}
\Pi_{\mathrm{good}}
=
|0\rangle\langle0|_{\mathrm{br}}
\otimes
|0^a\rangle\langle0^a|_{\mathrm{anc}}
\otimes
I_{\mathrm{data}},
\qquad
S_{\mathrm{good}}=I-2\Pi_{\mathrm{good}}.
\end{equation*}
This phase oracle flips only the component containing the desired affine output.
The reflection about the prepared state is
\begin{equation*}
S_{\Phi^{(j)}} = 2|\Phi^{(j)}\rangle\langle\Phi^{(j)}| - I,
\qquad
Q_j = -S_{\Phi^{(j)}}S_{\mathrm{good}}.
\end{equation*}
Writing
\begin{equation*}
|\Phi^{(j)}\rangle
=
\sqrt{p_j}\,|G_j\rangle+\sqrt{1-p_j}\,|B_j\rangle,
\qquad
\sin\theta_j=\sqrt{p_j},
\end{equation*}
after $m_j$ Grover iterations we obtain
\begin{equation*}
Q_j^{m_j}|\Phi^{(j)}\rangle
=
\sin\!\big((2m_j+1)\theta_j\big)|G_j\rangle
+
\cos\!\big((2m_j+1)\theta_j\big)|B_j\rangle.
\end{equation*}
Thus, by choosing $m_j\ge1$ appropriately, the desired data branch can be
amplified before the next affine step \cite{grover1,grover2}. The unscaled amplitudes may then be
recovered through post-processing together with additional overlap measurements.

\subsubsection{Cost Comparison with End-Stage Amplification}
Let $W_j$ denote the circuit that prepares the state after the first $j$ affine
steps, before any amplitude amplification is applied, and let $T(W_j)$ be its
gate cost. If $T_\ell$ is the cost of the $\ell$-th affine step, then
$
T(W_j)=\sum_{\ell=1}^{j} T_\ell,
\qquad
T(W_j)\approx j\bar T_W
$
when $\bar T_W$ is the average per-step cost.

In the end-stage strategy, after all $k$ affine steps the desired branch has
amplitude $1/2^k$, so its success probability is $p_k=1/4^k$. Standard
amplitude amplification therefore requires
$
m_{\mathrm{final}}=\Theta\!\left(\frac{1}{\sqrt{p_k}}\right)=\Theta(2^k)
$
Grover iterations. Since each iteration requires one forward and one inverse
application of the full $k$-step circuit, the total cost scales as
\[
C_{\mathrm{final}}
=
\Theta\!\bigl(2^k\,T(W_k)\bigr)
=
\Theta\!\bigl(2^k\,k\,\bar T_W\bigr).
\]

By contrast, in the interleaved strategy, after each affine step the desired
branch has amplitude $1/2$ and hence success probability $1/4$. Thus a
constant number of Grover iterations suffices at every stage (ideally one
iteration in the noiseless setting). The total amplification cost is then
\begin{equation*}
\resizebox{\columnwidth}{!}{$\displaystyle
C_{\mathrm{interleaved}}
=
\Theta\!\left(\sum_{j=1}^{k} T(W_j)\right)
\approx
\Theta\!\left(\sum_{j=1}^{k} j\bar T_W\right)
=
\Theta\!\bigl(k^2 \bar T_W\bigr).
$}
\end{equation*}

Therefore, interleaved amplitude amplification replaces the exponential
end-stage overhead by a polynomial one. In particular, the number of Grover
rounds drops from $\Theta(2^k)$ to $\Theta(k)$, while the total gate cost
improves from $\Theta(2^k k \bar T_W)$ to $\Theta(k^2 \bar T_W)$.



\begin{table}[!t]
\centering
\caption{Scaling comparison: end-stage vs.\ interleaved amplitude
amplification (AA) for $k$ sequential affine transformations.}
\label{tab:scaling_comparison}
\scriptsize
\begin{tabular}{@{}lcc@{}}
\toprule
\textbf{Metric} & \textbf{End-Stage AA} & \textbf{Interleaved AA} \\
\midrule
Target amplitude after $k$ steps & $1/2^k$ & $\approx 1$ \\
Success probability (pre-AA) & $1/4^k$ & $\approx 1$ \\
AA iterations & $O(2^k)$ & $k$ \\
Total unitary applications & $O(2^k \cdot k \cdot \bar T_W)$ & $O(k^2 \cdot \bar T_W)$ \\
Practical limit on $k$ & $\leq 10$ & $\leq 100{+}$ \\
\bottomrule
\end{tabular}
\end{table}

\begin{algorithm}[!t]
\scriptsize
\caption{Sequential Affine Transformations with Interleaved Amplitude
Amplification (AA)}
\label{alg:interleaved_aa}
\begin{algorithmic}[1]
\Function{InterleavedAffine}{$\Psi_0,\{A_1,\dots,A_k\},
\{B_1,\dots,B_k\},n$}
  \State $\ket{\phi} \gets \texttt{INITIALIZE\_AMPLITUDES}(\Psi_0, n)$
  \Comment{Prepare $n$-qubit input state}
  \State Allocate ancilla register $\ket{0^a}_{\mathrm{anc}}$
         and branching qubit $\ket{0}_{\mathrm{br}}$
  \For{$j = 1$ \textbf{to} $k$}
    \State \textbf{// --- Step $j$: Apply $j$-th affine transformation ---}
    \State $U_j \gets \texttt{BLOCK\_ENCODE}(A_j)$
    \Comment{Block encode $A_j$ into unitary $U_j$}
    \State $\tilde{B}_j \gets \texttt{RESCALE}(B_j, 2^{n})$
    \Comment{Rescale translation vector}
    \State $W_j \gets \texttt{COMPOSE}(U_j, \tilde{B}_j)$
    \Comment{Full step: block encoding $+$ Hadamard add/sub}
    \State $\ket{\Phi^{(j)}} \gets W_j \ket{\phi}$
    \Comment{Apply; target amplitude is $1/2$}
    \State
    \State \textbf{// --- Amplitude amplification (1 Grover iterate) ---}
    \State Construct $S_0 = I - 2\,\Pi_{\mathrm{good}}$
    \State Construct $S_{\mathrm{in}}^{(j)} =
           W_j(2\ket{0}\bra{0} - I)W_j^\dagger$
    \State $Q_j \gets -S_0 \cdot S_{\mathrm{in}}^{(j)}$
    \Comment{Grover iterate}
    \State $\ket{\phi} \gets Q_j\,\ket{\Phi^{(j)}}$
    \Comment{Amplify: amplitude $1/2 \to \approx 1$}
    \State
  \EndFor
  \State \Return $\ket{\phi}$
  \Comment{Contains $A_k(\cdots(A_1 \Psi_0 + B_1)\cdots) + B_k$}
\EndFunction
\end{algorithmic}
\end{algorithm}

\section{Applications}
\label{sec:apps}

\subsection{Discrete Signal Processing}

\begin{figure}[htbp]
    \centering
    \begin{minipage}[t]{0.4\textwidth}
        \centering
        \includegraphics[width=\linewidth]{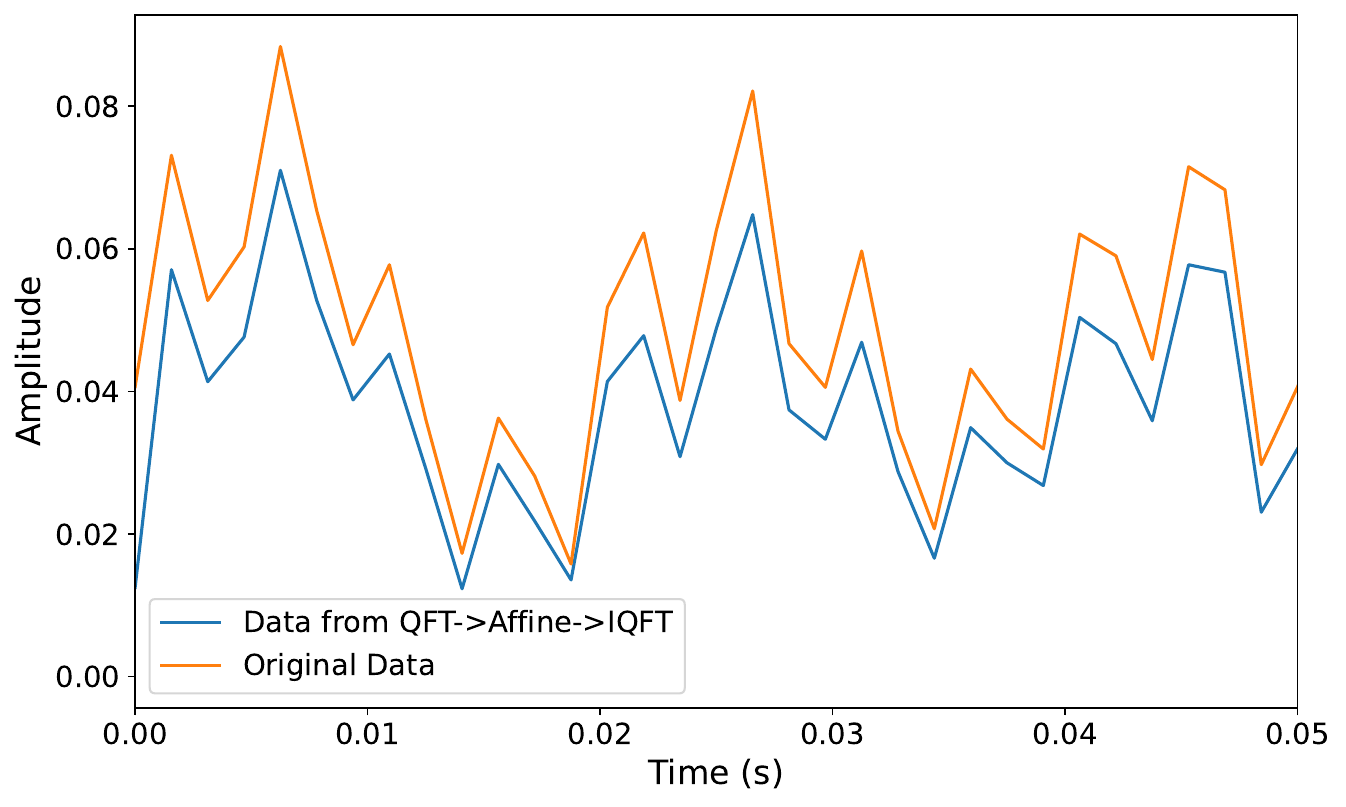}
        {\scriptsize (a) Illustrates the original signal and the result of our quantum procedure (QFT \(\rightarrow\) Affine Transformation \(\rightarrow\) Inverse QFT).}
    \end{minipage}
    \hfill
    \begin{minipage}[t]{0.4\textwidth}
        \centering
        \includegraphics[width=\linewidth]{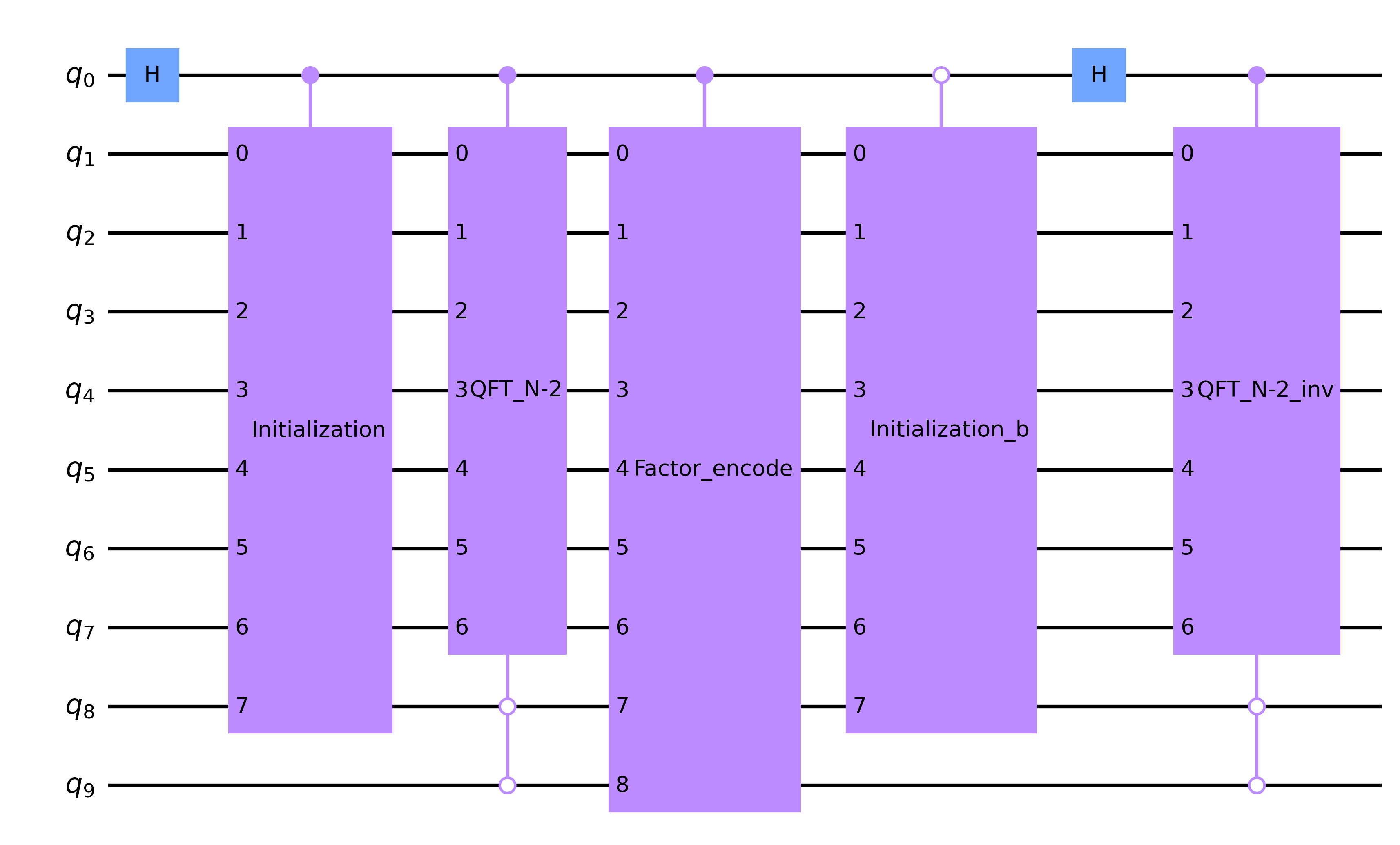}
        {\scriptsize (b) Schematic of a quantum circuit in Qiskit implementing these steps on a 10-qubit quantum system.}
    \end{minipage}
    \caption{\scriptsize Discrete signal processing using our quantum affine transformation algorithm.  Following a QFT operation on an input signal, a single affine transformation modifies a signal in Fourier space, and an inverse QFT transforms the altered signal back to the time domain.}
    \label{fig:signal_processing}
\end{figure}

Discrete signal processing techniques are pivotal in enhancing signals, suppressing noise, filtering image and audio data, and enabling efficient data compression~\cite{DSP1,DSP2}.  
These transformations facilitate precise control of signal properties in both the time and frequency domains.  
Classically, the discrete Fourier transform (DFT) of the signal vector \(X = [x_1, x_2, \ldots, x_N] \) is given by $
X_k = \sum_{n=1}^{N} x_n \, e^{-j \frac{2\pi}{N} (k-1)(n-1)}, \quad k = 1, 2, \ldots, N,$
where \( N \) is the signal length, and \( X_k \) represents the complex Fourier coefficients in the frequency domain~\cite{DSP3}. To manipulate the signal in the frequency domain, an affine transformation is applied to the Fourier coefficients $\tilde{X}_k = a \cdot X_k + b,$ where \( a \) is a scaling factor for amplitude modification, and \( b \) introduces a constant bias. The modified coefficients are transformed back into the time domain using the Inverse Discrete Fourier Transform (IDFT): $
\tilde{x}_n = \frac{1}{N} \sum_{k=1}^{N} \tilde{X}_k \, e^{j \frac{2\pi}{N} (k-1)(n-1)}, \quad n = 1, 2, \ldots, N.
$

We conduct an example of performing an affine transformation of a signal in frequency space using the quantum Fourier transform (QFT) and our proposed method.  
Given a discrete signal  
$x_n = \sum_{i=1}^{L} A_i \sin(2\pi f_i t_n)$,
where \( t_n \) represents uniform sampling in time, \( f_i \) are the frequencies, and \( A_i \) are the corresponding amplitudes of the sinusoidal components, we encode the \( A_i \) as amplitudes of a quantum state. We then perform QFT on the signal~\cite{DSP4}, followed by an affine transformation in the frequency domain (using our method), and then finally an inverse QFT. The results of this experiment on a random such input signal are shown in Fig.~\ref{fig:signal_processing}.


\subsection{Simulation of Driven Quantum Dynamics}
A driven linear quantum system is described by $
i\dot{\psi}(t)=H\psi(t)+\eta(t)$,
where $\psi(t)$ is the time-dependent state, $H$ is the Hamiltonian governing the internal homogeneous evolution, and $\eta(t)$ is an external driving or source term. Such equations arise in quantum mechanics as effective descriptions of externally driven systems, source-injected dynamics, or reduced equations obtained after eliminating auxiliary degrees of freedom. Closely
related inhomogeneous linear differential equations have also been studied in the context of quantum algorithms for differential equations, where the goal is to encode the solution of a driven or forced linear system into the amplitudes of a quantum state~\cite{berry1,berry2,berry3}.
For time-independent $H$ and initial state $\psi(0)=\psi_0$, the solution is:
\begin{equation*}
\psi(t)=e^{-iHt}\psi_0-i\int_0^t e^{-iH(t-s)}\eta(s)\,ds,
\end{equation*}
which is the Duhamel form of the solution. Discretizing time as $t_k=k\Delta t$ and writing $\psi_k:=\psi(t_k)$ gives the exact one-step relation:
\begin{equation*}
\psi_{k+1}=e^{-iH\Delta t}\psi_k-i\int_{t_k}^{t_{k+1}} e^{-iH(t_{k+1}-s)}\eta(s)\,ds.
\end{equation*}
Using a first-order approximation $\eta(s)\approx \eta_k:=\eta(t_k)$ over each interval yields the recursive update
$\psi_{k+1}\approx e^{-iH\Delta t}\psi_k-i\Delta t\,\eta_k$,
which has the affine form $\psi_{k+1}=U\psi_k+c_k$, with $U=e^{-iH\Delta t}$ and $c_k=-i\Delta t\,\eta_k$. Writing $U=e^{-iH\Delta t}$, the first three steps are:
\begin{align*}
\psi_1 &= U\psi_0+c_0,\\
\psi_2 &= U^2\psi_0+Uc_0+c_1,\\
\psi_3 &= U^3\psi_0+U^2c_0+Uc_1+c_2.
\end{align*}
The quantum circuit that implements the desired state evolution to get $\psi_3$ is given in Fig.~\ref{fig:driven_affine_circuit}. For a larger number of iterations, interleaved amplitude amplification can be inserted at the locations indicated by the vertical dashed lines in the circuit.

\begin{figure}[t]
\resizebox{\columnwidth}{!}{%
\begin{quantikz}[
row sep={0.4cm,between origins},
column sep=0.06cm
]
\lstick{$q_1$}
& \gate{\scriptstyle H}
& \octrl{1}
& \octrl{2}
& \octrl{3}
& \octrl{3}
& \octrl{3}
& \octrl{3}
& \octrl{3}
& \octrl{3}
& \octrl{3}\slice{}      
& \octrl{2}
& \octrl{2}
& \octrl{3}\slice{}      
& \octrl{1}
& \ctrl{1}
& \gate{\scriptstyle H}\slice{}   
& \qw
\\
\lstick{$q_2$}
& \qw
& \gate{\scriptstyle H}
& \octrl{1}
& \octrl{2}
& \octrl{2}
& \octrl{2}
& \octrl{2}
& \octrl{2}
& \octrl{2}
& \octrl{2}
& \octrl{1}
& \ctrl{1}
& \ctrl{2}
& \gate{\scriptstyle H}
& \gate[5]{\scriptstyle c_2}
& \qw
& \qw
\\
\lstick{$q_3$}
& \qw
& \qw
& \gate{\scriptstyle H}
& \octrl{1}
& \octrl{1}
& \octrl{1}
& \octrl{1}
& \ctrl{1}
& \ctrl{1}
& \ctrl{1}
& \gate{\scriptstyle H}
& \gate[4]{\scriptstyle c_1}
& \qw
& \qw
& \ghost{\scriptstyle c_2}
& \qw
& \qw
\\
\lstick{$q_4$}
& \qw
& \qw
& \qw
& \gate[3]{\scriptstyle \psi_0}
& \gate[3]{\scriptstyle U}
& \gate[3]{\scriptstyle U}
& \gate[3]{\scriptstyle U}
& \gate[3]{\scriptstyle c_0}
& \gate[3]{\scriptstyle U}
& \gate[3]{\scriptstyle U}
& \qw
& \ghost{\scriptstyle c_1}
& \gate[3]{\scriptstyle U}
& \qw
& \ghost{\scriptstyle c_2}
& \qw
& \qw
\\
{}
& \vdots
& \vdots
& \vdots
& \ghost{\scriptstyle c_0}
& \ghost{\scriptstyle U}
& \ghost{\scriptstyle U}
& \ghost{\scriptstyle U}
& \ghost{\scriptstyle c_0}
& \ghost{\scriptstyle U}
& \ghost{\scriptstyle U}
& \vdots
& \ghost{\scriptstyle c_1}
& \ghost{\scriptstyle U}
& \vdots
& \ghost{\scriptstyle c_2}
& \vdots
&
\\
\lstick{$q_n$}
& \qw
& \qw
& \qw
& \ghost{\scriptstyle c_0}
& \ghost{\scriptstyle U}
& \ghost{\scriptstyle U}
& \ghost{\scriptstyle U}
& \ghost{\scriptstyle c_0}
& \ghost{\scriptstyle U}
& \ghost{\scriptstyle U}
& \qw
& \ghost{\scriptstyle c_1}
& \ghost{\scriptstyle U}
& \qw
& \ghost{\scriptstyle c_2}
& \qw
& \qw
\end{quantikz}
}
\caption{Quantum circuit illustrating three sequential affine-transformation steps. Open and closed controls indicate conditioning on \(\ket{0}\) and \(\ket{1}\), respectively. The register from \(q_4\) to \(q_n\) denotes the data subspace, with vertical dots indicating possible intermediate qubits. Blocks labeled \(\psi_0\), $c_0$, $c_1$, and $c_2$ are conditional state-preparation operations and blocks labeled \(U\) are conditional unitaries. Red dash lines represent the stage where we can perform single step Amplitude Amplification.}
\label{fig:driven_affine_circuit}
\end{figure}

\section{Conclusion}
In summary, this work presents a novel technique for applying sequences of affine transformations to the probability amplitudes of an input quantum state.
By encoding a non-unitary matrix into a unitary and using Hadamard supported conditional initialization, a single step affine transformation is applied, and this process is generalized to any number of steps. The bottleneck of amplitude decay for larger number of steps is solved by interleaved single step amplitude amplification. Preliminary demonstrations of running our algorithm on a discrete signal processing task and on a driven dynamics evolution problem validate our implementation.

As quantum computing continues to advance, kernels like sequential affine transformations will be increasingly valuable for tackling complex computational tasks spanning optimization, machine learning, signal processing, differential equations, and beyond.
Future research could extend our approach to explore its potential in quantum machine learning and combinatorial optimization applications.

\clearpage

\bibliographystyle{ieeetr}
\bibliography{ref}

\end{document}